
\input phyzzx
\overfullrule=0pt
\def\n{\noindent}
\def\p{\partial}
\def\ge{\varepsilon}

\def\gl{\lambda^2}

\def\sk{\sqrt{\kappa} }
\def\IN{\relax{\rm I\kern-.18em N}}
\font\captura=cmr9
\null
\rightline {UTTG-22-93}
\rightline {July 1993}

\title{A Theory of Quantum Black Holes: Non-Perturbative
Corrections and No-Veil Conjecture
\foot{Work supported in part by NSF grant PHY 9009850 and
R.~A.~Welch Foundation.}}

\author{Jorge G. Russo }
\address {Theory Group, Department of Physics, University of
Texas\break
Austin, TX 78712}

\abstract
A common belief is that further
quantum corrections near the singularity of a large black hole
should not substantially modify the semiclassical picture of black
hole evaporation; in particular, the outgoing spectrum of radiation should
be very close to the thermal spectrum predicted by Hawking.
In this paper we explore a possible counterexample: in the context of
dilaton gravity, we find that non-perturbative quantum corrections
which are important in strong coupling regions may completely alter the
semiclassical picture, to the extent that the presumptive space-like boundary
becomes time-like, changing in this way the causal structure of the
semiclassical geometry. As a result, only a small fraction of the total
energy is radiated outside the fake event horizon; most of the energy
comes in fact at later retarded times and there is no
information loss problem.
Thus we propose that this may constitute a general characteristic of quantum
black holes, that is, quantum gravity might be such as to prevent the
formation of global event horizons. We argue that this is not
unnatural from
the viewpoint of quantum mechanics.

\endpage

\Ref\hawi{S.~W.~ Hawking,
Comm. Math. Phys.  43 (75) 199.}

\Ref\hawii{S.~W.~Hawking,
Phys. Rev.  D14 (76) 2460.}

\Ref\bps{D.N. Page, Phys. Rev. Lett.44 (1980) 301;
T.~Banks, M.~Peskin and L.~Susskind, Nucl. Phys. B244 (84) 125.}

\Ref\thoof{G. 't Hooft, Nucl. Phys. B335 (1990) 138.}

\Ref\stu {L. Susskind, L. Thorlacius and J. Uglum, Stanford  preprint,
SU-ITP-93-15 (1993).}

\Ref\thooft {G. 't Hooft, talk at Santa Barbara Conference
``Quantum aspects of black holes", June 1993.}

\Ref\endpoint{J.~G.~Russo, L.~Susskind and L.~Thorlacius,
 Phys. Rev. D46 (1992) 3444.}

\Ref\cosmic {J.~G.~Russo, L.~Susskind and L.~Thorlacius,
 Phys. Rev. D47 (1993) 533. }

\Ref\cghs{C.~G.~Callan, S.~B.~Giddings, J.~A.~Harvey and
A.~Strominger, Phys. Rev.  D45 (1992) R1005.}

\Ref\rt{J.G.~Russo and A.A.~Tseytlin, Nucl. Phys.~B382 (1992) 259.}

\Ref\banks{T. Banks, A. Dabholkar, M.R. Douglas and M. O'Loughlin,
Phys.Rev.D45 (1992) 3607.}

\Ref\rst{J.~G.~Russo, L.~Susskind and L.~Thorlacius,
Phys. Lett. B292 (1992) 13.}

\Ref\bghs{B.~Birnir, S.~B.~Giddings, J.~A.~Harvey and A.~Strominger,
 Phys. Rev.  D46 (92) 638;
S.~W.~Hawking, Phys. Rev. Lett.  69 (92) 406;
L.~Susskind and L.~Thorlacius, Nucl. Phys.~ B382 (1992) 123.}

\Ref\andy{A.~Strominger, Phys.Rev.D46 (1992) 4396.}

\Ref\bilcal{A.~Bilal and C.~G.~Callan, Nucl. Phys. B394 (1993) 73.
 S.~P.~de~Alwis, Phys.Lett.B289 (1992) 278; Phys.Lett.B300 (1993) 330.}

\Ref\gidstr{S.~B.~Giddings and A.~Strominger, Phys. Rev. D47 (1993) 2454.}

\Ref\gidnel{S.~B.~Giddings and W. Nelson, Phys.Rev. D46 (1992) 2486.}

\Ref\hawste{D. Lowe, Phys.Rev. D47 (1993) 2446;
S.~W.~Hawking and J.~M.~Stewart, University of Cambridge
preprint, PRINT-92-0362, 1992.}

\Ref\positive{ Y. Park and A. Strominger, Phys. Rev. D47 (1993) 1569;
A. Bilal, Princeton preprint, PUPT-1373 (1993).}

\Ref\banka{T. Banks, M. O'Loughlin and A. Strominger, Phys. Rev. D47
(1993) 4476.}

\Ref\verlinde{E. Verlinde and H. Verlinde, Princeton preprint, PUPT-1380,
IASSNS-HEP-93/8, 1993; K. Schoutens, E. Verlinde and H. Verlinde,
Princeton preprint, PUPT-1395, IASSNS-HEP-93/25, 1993.}

\Ref\stephen{S.W. Hawking and J.D. Hayward, University of Cambridge and
CalTech preprint, DAMTP-R93-12, CALT-68-1861.}

\Ref\dpage{D.N. Page, University of Alberta preprint, Alberta-Thy-23-93,
(1993).}

\Ref\andywhite{A. Strominger, Santa Barbara preprint, NSF-ITP-93-92
(1993).}

\Ref\ddk{F. David, Mod. Phys. Lett. A3 (1988) 1651;
J. Distler and H. Kawai, Nucl. Phys. B321 (1989) 509.}

\Ref\witten{E.~Witten, Phys. Rev.  D44 (1991) 314.}

\Ref\wald{R. M. Wald, {\it General Relativity}, (University of
Chicago Press, Chicago, 1984).}

\chapter{Introduction}

The proposal of Hawking that black holes evaporate by emitting
thermal radiation [\hawi ] led to a puzzling and confusing status about the
fate of quantum information in gravitational collapse. In particular,
this proposal entails allowing pure states to evolve into mixed states,
requiring a modified version of quantum mechanics in order to accomodate
loss of quantum coherence [\hawii ]. This proposal was received with some
criticism by different authors (see e.g. ref. [\bps ]),
but so far an understanding of
the phenomenon has not been achieved. Apart from the existence of
a mathematical framework which may
reconcile Hawking observation with quantum mechanics, information loss
raises a serious question of principle for observers who do not fall
into black holes, their future being uncorrelated to their
past. Several attempts have been made to provide alternative ways
before accepting that quantum mechanical information is simply
lost in the process of black hole evaporation. Thus far the proposals
fall either into one of the two following categories:
1) by the end of the evaporation process there is a planckian-size
stable or long-lived remnant that still contains the information;
2) the back reaction
to the emission of radiation and quantum corrections
introduce subtle correlations
between different modes, allowing the information
to come out continuously encoded in the Hawking radiation, the process
being described by a unitary $S$-matrix.

These two approaches are not exempt from criticism. The
first has problems with CPT and also with thermodynamics.
The second possibility seems to imply acausal propagation of the
information, since this was carried far beyond the horizon before the
curvature is strong enough for quantum gravitational effects to be important.
To avoid the serious problem of acausality,
a rather temerarious proposal that has recently been revived [\thoof -\thooft ]
consists in postulating that the information is duplicated at the moment it
crosses the horizon. This interpretation requires planckian physics
occurring in the vicinity of the horizon, and implies a dual description
of reality at macroscopic levels. A paradigm is the following:
a couple of astronauts in
an inertial rocket falling into a large black hole do not feel anything
special at the moment they cross the horizon; they
have  long and prolific
lives inside the black hole, they have sons, daughters, etc.
But in the view of an outside observer rocket and passengers will have been
completely burned
out at the moment they touch a very hot, planckian temperature surface, namely
the event horizon.
The outside
observer recovers every bit of information of the rocket and detects that
the astronauts died when they were a few Planck units away from the
event horizon. Is this an obvious contradiction?
The advocates of this proposal argue that there is
no logical contradiction because the free-falling observers are unable to
communicate with the outside observer.

In this paper we present a perhaps more conservative possibility which
does not belong to the schemes 1) and 2) mentioned above.
Fig. 1a is a Penrose diagram representing the standard picture of
semiclassical black hole evaporation, which is reliable in the region
away from the singularity. Fig. 1b completely agrees with
fig. 1a in all regions where the semiclassical equations of motion
are suppossed to apply, but instead of a singularity there is simply
a strong curvature region, and the actual boundary of the space-time is
time-like. This requires boundary conditions. Let us assume
that some sort of reflecting boundary conditions can be imposed there and,
maybe, they lead to a finite curvature on the boundary, just as it happens
in the low-energy sector of the two-dimensional model of ref.
[\endpoint, \cosmic], that we shall review in sect. 2.
The causal structure of fig. 1b is completely different than the causal
structure of fig. 1a, and therefore one would expect that the corresponding
spectra of outgoing Hawking radiation should be distinct, perhaps in a
crucial way.
In fact, this turns out to be the case. Given a geometry like fig.1b, with
reflecting-type boundary conditions on the time-like boundary,
most of the energy shall appear in the region
in causal contact with the time-like boundary, i.e. far beyond the
fake event horizon, and thus there is no information loss problem.
The resulting picture is in some sense similar to the low-energy sector
of ref. [\endpoint, \cosmic], and it does not differ much from an
accelerating mirror.
The boundary of space-time actually being time-like, there is no longer
any reason to believe
that an unitary $S$-matrix for the model cannot be constructed.
There are other similar scenarios that will be mentioned
in sect. 6, all of them leading essentially to the same conclusion.

\bigskip

\chapter{Semi-classical dilaton gravity}

A simplified model for black hole formation and
evaporation, known as the CGHS model, was introduced in ref. [\cghs ].
This model permits to study the Hawking phenomenon in detail,
avoiding all the mathematical complications of higher-dimensional
theories.
Different discussions on two-dimensional dilaton gravity can be
found, e.g., in refs. [\endpoint -\stephen ].
Let us consider the model introduced in ref. [\endpoint ].
In the conformal
gauge $g_{++}=g_{--}=0$, $g_{+-}=-{1\over 2} e^{2\rho}$, the effective
action containing the conformal anomaly can be written as
$$S={1\over \pi}\int d^2x \big[ -\p_+\chi \p_-\chi +
\p_+\Omega \p_-\Omega +\gl e^{(2/\sqrt{\kappa })(\chi-\Omega )}
+{1\over 2}\sum_{i=1}^N\p_+f_i\p_-f_i\big]\ ,
\eqn\action
$$
where $f_i$ are $N$ conformal fields, $\kappa=(N-24)/12 >0$ and
$$
\Omega={\sqrt{\kappa}\over 2}\phi +{e^{-2\phi}\over\sqrt\kappa }\ ,
\ \ \ \chi -\Omega =\sk (\rho-\phi )\ \ .
\eqn\fields
$$
The constraints are
$$
\kappa t_{\pm}(x^\pm )=-\p_\pm\chi \p_\pm\chi + \p_\pm\Omega \p_\pm\Omega
+\sqrt{\kappa}\p^2_\pm\chi +{1\over 2}\sum_{i=1}^N\p_\pm f_i\p_\pm f_i\ .
\eqn\constraints
$$
The functions $t_\pm (x^\pm )$ reflect the nonlocal nature of the anomaly
and are determined by boundary conditions.
The solution to the semi-classical equations
of motion and the constraints,
for general distributions of incoming matter, is given  in Kruskal
coordinates by
$$
\Omega=\chi=-{\gl\over\sk }x^+\big( x^-+{1\over \gl }P_+(x^+)\big)
+{M(x^+)\over \sk\lambda } -{\sk \over 4}\ln (-\gl x^+x^-)\ ,
\eqn\solutions
$$
where $M(x^+)$ and $P_+(x^+)$ respectively represent total energy and
Kruskal momentum of the incoming matter
at advanced time $x^+$:
$$
M(x^+)=\lambda \int _0^{x^+} dx^+ x^+ T_{++}(x^+)\ ,
\ \ P_+(x^+)= \int _0^{x^+} dx^+ T_{++}(x^+)\ \ .
\eqn\enermom
$$
In the case $T_{++}=0$ one obtains the familiar linear dilaton vacuum,
$e^{-2\phi}=e^{-2\rho}=-\gl x^+x^-$.

Generically, there will be a curvature singularity at
$\phi=\phi_{\rm cr}=-{1\over 2}\ln (\kappa /4)$, which can be regarded
as the boundary of the space-time.

Let us assume that originally the geometry is the linear dilaton vacuum
and at some time, which we arbitrarily set at $x^+=1/\lambda $,
the incoming flux is turned on.
As observed in ref. [\endpoint ],
there are two different regimes, according to
whether the incoming matter energy-momentum tensor is less or
greater than a critical flux
$$
T_{++}^{\rm cr}(x^+)={\kappa\over 4} {1\over {x^+}{}^2}\ .
\eqn\critflux
$$

In the supercritical regime the line $\phi=\phi_{\rm cr}$ is space-like and one
has a time-depending geometry representing the process of formation and
evaporation of
a black hole (see fig. 2). At the endpoint line, $x^-=x^-_s, \ x^+>x^+_s$,
it is possible to match the solution continuously with the linear dilaton
vacuum.

In the subcritical regime the boundary is time-like and
one needs boundary conditions in order to determine the evolution in the
region in causal contact with the time-like boundary (see fig. 3).
It turns out that there are natural, reflecting-type boundary conditions
which uniquely determine the evolution and implement cosmic
censorship hypothesis [\cosmic ]; they are in fact the only possible
boundary conditions which lead to a finite curvature on
the boundary line.

The curvature scalar is $R=8e^{-2\rho}\p_+\p_-\rho$,
$$
\p_+\p_-\rho ={1\over\Omega'}\bigg[ \p_+\p_-\chi -
{4 e^{-2\phi}\over \sk} \p_+\phi \p_-\phi \bigg ]\ .
\eqn\curvature
$$
At $\phi=\phi_{\rm cr}$ one has $\Omega '(\phi )=0$. In the Kruskal gauge,
$\p_+\p_-\chi =-{\gl\over\sk }$. Therefore, in order for the curvature
to be finite at $\phi=\phi_{\rm cr}$, it is necessary that
$$
\p_+\phi \p_-\phi =-{\gl\over\kappa}\ .
$$
In particular, this implies (cf. eq. \fields )
$$
\p_+\Omega \big|_{\phi=\phi_{\rm cr} }=
\p_-\Omega \big|_{\phi=\phi_{\rm cr} }=0
\eqn\bc
$$
As a result, the solution in region (ii) is given by
$$
\Omega ^{\rm (ii)}(x^+,x^-)=\Omega ^{\rm (i)}(x^+,x^-)+F(x^-)
\eqn\omegaii
$$
where $\Omega ^{\rm (i)}$ is given by eq. \solutions ,
$$
F(x^-)={\sk \over 4}\ln (-\gl x^-\hat x^+)-{M(\hat x^+)\over \sk \lambda}
-{\sk\over 4}\ln ({\kappa\over 4})\ ,
\eqn\fff
$$
and $\hat x^+(x^-)$ is the boundary curve given by
$$
{\kappa\over 4}=-\gl \hat x^+ \big( x^-+{1\over \gl }P_+(\hat x^+)\big)\ .
\eqn\bcurve
$$

In ref. [\cosmic ] it was shown that these boundary conditions conserve energy.
Let $m\equiv M(\infty )$ and $p\equiv P_+(\infty )$.
The outgoing energy fluxes $T_{--}(x^-)$
in region (i) and (ii) are, respectively
$$
T^{\rm (i)}_{--}(x^-)={\kappa\over 4}\big[ {1\over (x^-+{1\over\gl }p)^2}
- {1\over {x^-}^2} \big]\ ,
\eqn\tini
$$
$$
T^{\rm (ii)}_{--}(x^-)={\kappa\over 4} {1\over (x^-+{1\over\gl }p)^2}
-{\lambda^4 \over {\kappa\over 4 \hat x^+}- \hat x^+ T_{++}(\hat x^+) }\ .
\eqn\tinii
$$
Note that $T^{\rm (i)}_{--}(x^-)\sim 0$ for $x^-<<x^-_1=-p/\gl$.

The total radiated energies in region (i) and region (ii) are
$$\eqalign{
E_{\rm out}^{\rm (i)}&=-\lambda \int_{-\infty}^{x_0^-} dx^-
\big( x^-+{1\over\gl }p\big) T_{--}^{\rm (i)}  \cr
&=p -{\kappa\lambda\over 4}\ln \big( 1- {4p\over \kappa\lambda}\big)\ ,
\cr}
\eqn\eneri
$$
$$\eqalign{
E_{\rm out}^{\rm (ii)}&=-\lambda \int^{x^-_1}_{x_0^-} dx^-
\big( x^-+{1\over\gl }p\big) T_{--}^{\rm (ii)} \cr
&=m-p
+{\kappa\lambda\over 4}\ln \big( 1- {4p\over \kappa\lambda}\big)\ .
\cr}
\eqn\enerii
$$

A close examination to eqs. \tini - \enerii\ shows that
for low energy fluxes one has
$$
E_{\rm out}^{\rm (i)}<<m\ ,\ \ \ E_{\rm out}^{\rm (ii)}\sim m \ ,
\eqn\salepoco
$$
that is, most of the energy comes out by pure reflection on the
space-time boundary.

Instead, for energy fluxes near the critical value,
one may have $E_{\rm out}^{\rm (i)}>m$ and $E_{\rm out}^{\rm (ii)}<0$.
This implies that the semiclassical approximation is breaking down
some Planck units before entering into region (ii).
Perhaps non-perturbative contributions are already important.

\bigskip

\chapter{Non-perturbative quantum corrections }

The form of the quantum effective action which includes the quantum
anomaly term of exact semiclassical dilaton gravity [\rt, \bilcal,
\gidstr , \endpoint ] follows by a DDK-type argument [\ddk].
Instead of using an invariant regularisation (which is complicated in the
conformal gauge) one adopts a non-invariant cut-off adding at the same time
some counterterms which are necessary in order to satisfy the reparametrisation
invariant Ward identities. The resulting `effective action' should generate
a theory which is invariant under the background Weyl symmetry,
$\hat g\to e^{2\tau(x)}\hat g,\ \rho(x)\to\rho(x)-\tau (x)$, where $\hat g$
is a background metric. Since the metric
$g=e^{2\rho}\hat g$ is left unchanged this transformation should be an exact
symmetry of the theory, i.e. the $\beta $- funcitons of the couplings in the
`effective action' should vanish. The basic assumption is that the
conformal factor dependence of the covariant quantum measure and regularisation
can be represented by a local effective action containing only simplest
lowest derivative terms. The kinetic term of the resulting quantum effective
action is
modified by the Weyl anomaly term and also by possible counterterms
and one automatically attains a partial resummation of the standard loop
expansion. In particular, this procedure also generates counterterms
which are of non-perturbative nature, e.g. of the form $e^{-1/g^2}=\exp
[-e^{-2\phi}]$.

Thus one considers the two-dimensional sigma-model with fields $\rho $
and $\phi $ and fix the couplings by demanding conformal invariance.
Since the target metric is flat it is possible to go
to the diagonal parametrization $\chi $ and $\Omega $ which simplifies
the equations of motion, as found in ref. [\bilcal ].
 To leading order, the $\beta $-function corresponding to the `tachyonic'
coupling is given by
$$
\beta^V=\big[ {\p^2\over \p \chi^2 }-{\p^2\over \p \Omega ^2 }
+{1\over 2}\sk {\p\over \p \chi }-1 \big ]V(\chi,\Omega )\ .
\eqn\betav
$$
The potential $\gl e^{(2/\sqrt{\kappa })(\chi-\Omega )}$
employed in sect. 2 (see eq. \action )
is, in fact, a particular solution to this equation.
The most general solution is
$$
V(\chi,\Omega )=\int da \big[ v^+(a) e^{a\chi+b\Omega }+
v^-(a) e^{a\chi-b\Omega }\big]\ ,
\eqn\pot
$$
with
$$
b=- \sqrt {a^2+{\sk\over 2}a-1}\ .
\eqn\bbb
$$
However, we are interested in solutions which in the weak coupling region
$e^\phi\to 0$ lead to the classical CGHS action. From eq. \fields\
we deduce that this requires the condition
$$
a+b\leq 0\ .
$$
Thus the general potential which leads to the CGHS action in
the weak coupling regime is
$$
V(\chi,\Omega )=\gl e^{(2/\sqrt{\kappa })(\chi-\Omega )}
+\int _{2/\sk }^\infty da\  v(a) e^{a\chi+b\Omega }\ .
\eqn\poten
$$
The second term will contain non-perturbative contributions of the form
$\exp [{(a+b)\over \sk} e^{-2\phi}]$.
At full quantum level one expects that all physical quantities
should be plagued of non-perturbative corrections, originating from
strong curvature regions.

The semiclassical equations of motion become complicated when
the general potential \poten\ is adopted.
In order to ellucidate the basic idea,
let us consider a simple model with the action given by
$$
S={1\over \pi}\int d^2x \big[ -\p_+\chi \p_-\chi +
\p_+\Omega \p_-\Omega +\gl e^{(2/\sqrt{\kappa })(\chi-\Omega )}
+\mu e^{a\chi+b\Omega}
+{1\over 2}\sum_{i=1}^N\p_+f_i\p_-f_i\big]\ ,
\eqn\npaction
$$
where $\Omega$ and $\chi $ are given in terms of $\phi$ and $\rho$
as in eq. \fields  , $\mu$ is for the moment  arbitrary,
$a>2/\sk $ and $b$ is given by eq. \bbb . The RST model is
a particular case with $\mu=0 $.

In principle, non-perturbative corrections may be different for
distinct geometries. In particular, it may be that the linear dilaton vacuum
does not receive any correction at all, the curvature being zero everywhere.
Unfortunately, a systematic derivation of the DDK ansatz --i.e. calculating
loops with covariant regularization, etc.--
is lacking, so it is unclear what should be the value of $\mu $ in
the ``phenomenological" action \npaction .
The only independent constants of the theory are $\lambda $ and $\kappa $.
Therefore
$\mu $ should be given in terms of $\lambda  $, $\kappa $ and maybe other
parameters characterizing the geometry, the ADM mass, {\it verbigratia},
or the moments
$$
P^n_+=\int_0^{\infty }dx^+ (x^+)^{-n+1}T_{++}(x^+)\ ,\ \ \ n\in \IN \ .
\eqn\momenti
$$
In the particular case of a shock-wave geometry, which will be investigated
in detail below, $\mu $ would only depend on $\lambda, \kappa$ and the
total ADM energy carried by the collapsing matter.

The essential point, i.e. that non-perturbative corrections can change
the space-time topology by turning the boundary curve from space-like
to time-like, is a quite generic result, holding true for a wide
range of choices for $\mu $.

The equations of motion are
$$
\p_+\p_- X=A e^Y\ \ ,
\eqn\eqx
$$
$$
\p_+\p_- Y=B e^X +Ce^Y\ \ ,
\eqn\eqy
$$
where
$$
X={2\over \sk}(\chi-\Omega )\ ,\ \ \ Y=a\chi+b\Omega \ ,
\eqn\xxyy
$$
and
$$
A=-{\mu\over \sk}(a+b) \ ,\ \ \ B=-{\gl\over \sk }(a+b)\ ,\ \ \
C={\mu\over 2} (b^2-a^2)\ .
\eqn\abc
$$

A particular solution $X=Y+$const. is easily found, for which the system
reduces to the Liouville equation. Unfortunately,
this solution does not satisfy
the asymptotic conditions corresponding to black-hole configurations, and
thus it is uninteresting for our purposes.

Although the general solution to the above system of non-linear, partial
differential equations has not been written in closed form, to
our knowledge, it is nevertheless possible to obtain some exact, interesting
results, as we shall see below.

To leading order in an expansion
in powers of $\ge=\exp [{(a+b)\over \sk} e^{-2\phi}]$, the general solution
can be explicitly found by direct integration.
Let us consider the case of a shock-wave geometry representing
an infalling shell of matter
 by patching together a vacuum
configuration on the inside and, on the outside, a solution which
asymptotically corresponds to a black hole.
Let us write
$$X=X_{\rm sc}+O(\ge )\ ,\ \ \ Y=Y_{\rm sc}+O(\ge )\ ,
\eqn\nidea
$$
where, in  Kruskal coordinates,
$$X_{\rm sc}=0\ ,\ \ \
\eqn\tampoco
$$
$$Y_{\rm sc}=(a+b)[{m\over\lambda\sk }-{\gl\over\sk}x^+(x^-+{m\over\gl })]
 -(a+b){\sk\over 4}\ln (-\gl x^+x^-) \ ,\ \ \ x^+\geq 1/\lambda \ .
\eqn\semicla
$$
Now let us pick some convenient value for $a$ which will simplify the
calculation. For $\kappa >16$ we can
choose $a+b=-4/\sk $, i.e.
$$
a= {2\over \sk} {\kappa +16\over \kappa-16}\ .
\eqn\aaa
$$
For this value of $a$ we have
$$
e^{Y_{\rm sc}}
=-\gl x^+ x^-  \exp\big[ {4\gl \over \kappa} x^+(x^-+{m\over\gl })
-{4m\over \kappa \lambda}\big] \ ,
\eqn\eysc
$$
and
$$
A={4\mu\over\kappa}\ ,\ \ \ B={4\gl\over \kappa}\ ,\ \ \ C={16\mu\over
\kappa-16}\ .
\eqn\aabbcc
$$
The equations of motion take the form
$$
\p_+\p_- X=A e^{Y_{\rm sc}} +O(\ge ^2)\ ,
\eqn\eqxx
$$
$$
\p_+\p_- Y=B(1+X)+Ce^{ Y_{\rm sc}} +O(\ge ^2)\ .
\eqn\eqyy
$$
These equations can now be solved by direct integration.
The solution is uniquely determined by the boundary conditions
that at $x^-<<0 $ it must approach the semiclassical solution, eqs.
\tampoco , \semicla ,
and, on the infalling line $x^+=1/\lambda$, it must reduce to the
linear dilaton vacuum. We find
$$
X={\mu\kappa\over 4\gl}
\bigg[ e^{-{4m\over \lambda\kappa }}{\rm Ei}(-r) -
e^{-{4m\over \lambda\kappa }} {\rm Ei}\big(
{4\lambda\over \kappa }(x^-+{m\over\gl})\big)
-{x^-\over x^-+{m\over\gl }}\big( e^{-{4m\over \lambda\kappa }}
e^{-r}-e^{{4\lambda\over\kappa}x^-}\big)\bigg]\ ,
\eqn\xsol
$$
$$
Y=-r -{4m\over\kappa\lambda} +\ln (-\gl x^+x^-)
+y(x^+,x^-)-y({1\over\lambda},x^-)\ ,
\eqn\ysol
$$
where
$$
y(x^+,x^-)=\mu e^{-{4m\over \lambda\kappa }}x^+\big( x^-+{2m\over\gl }\big)
\big[{\rm Ei}(-r) -{\rm Ei}\big({4\lambda\over \kappa }(x^-+{m\over\gl})\big)
\big]-
$$
$$
{\kappa \mu \over 4\gl }e^{-{4m\over \lambda\kappa }}\big[
2 {\rm Ei}(-r) +  {\gl x^-+2m\over \gl x^-+m}e^{-r} \big] +
$$
$$
{\kappa^2\mu\over (\kappa-16)\gl }e^{-{4m\over \lambda\kappa }}
\bigg[ {\rm Ei}(-r) -{x^-\over x^-+{m\over\gl} }e^{-r}\bigg]
+{\kappa\mu\over 2\lambda}x^+e^{{4\lambda\over\kappa}x^-}\ ,
\eqn\unos
$$
$$
r\equiv -{4\gl\over\kappa }x^+(x^-+{m\over\gl })
\ ,\ \ \
{\rm Ei}(-r)=\int_\infty^r du {e^{-u}\over u} \ .
\eqn\dos
$$
Let us first consider $\kappa \sim 16$. In this case some terms can be
ignored, which renders the analysis simpler.
We obtain
$$
\chi=\Omega=-{\gl\over 4}x^+(x^-+{m\over\gl })+{m\over 4\lambda}
-\ln (-\gl x^+x^-) +$$
$${128\mu\over \gl(\kappa -16)} \bigg[ {x^-\over x^-+{m\over\gl }}
\big( e^{-{m\over 4\lambda }}
e^{-r}-e^{{\lambda\over 4}x^-}\big) -e^{-{m\over 4\lambda }}{\rm Ei}(-r)
+e^{-{m\over 4\lambda }}{\rm Ei}\big( {1\over 4}(x^-+{m\over\gl})\big)\bigg]\ .
\eqn\omga
$$
The points where $\p_+ \Omega =0$ indicate the position of the apparent
horizon (for a review of apparent horizon in 1+1 dimensions see appendix).
The apparent horizon is the border where the curves of constant $\phi $
change from time-like to space-like and conversely.
{}From eq. \omga\  one obtains
$$
\p_+\Omega=-{\gl \over 4}(x^-+{m\over\gl }) -{1\over x^+}+
{32 \mu\over \kappa -16}  e^{-{m\over 4\lambda}} e^{-r}\big( x^--
{4\over \gl x^+}\big)\ .
\eqn\plusomga
$$
The line $\phi=\phi_{\rm cr}$ intersects $x^+=1/\lambda $
at $x^-_0=-{\kappa\over 4\lambda}$. An inspection to eq. \plusomga \
reveals that the equation $\p_+\Omega=0$ may admit more than one
solution in the physical region $x^-<x^-_0 $
of the line $x^+=1/\lambda $.
In particular, if $\mu $ happens to obey
$$
\mu < \hat \mu (\lambda,\kappa, m)\ ,
\eqn\nombre
$$
with
$$
\hat \mu(\lambda,\kappa \sim 16, m)\cong - {e(\kappa-16)\over 1024}
m\lambda\ ,
\eqn\quemu
$$
then, for black holes with $m>>{\kappa \over 4}\lambda $,
there will be two apparent horizons in the region $x^-<x^-_0$; one at
$x^-_1\cong -{m\over\gl }-{\kappa\over 4\lambda}$
and the other at some $x^-_2$ near $-{\kappa\over 4\lambda}$.
The lines of constant $\phi $ will be time-like for
$x^-<x^-_1$, space-like for $x^-_1<x^-<x^-_2$, and again time-like
for $x^->x^-_2$.
In particular, this means that the boundary curve $\phi=\phi_{\rm cr}$
will start being time-like and therefore boundary conditions will be
necessary in order to determine the evolution in the region in
causal contact with the time-like boundary.

In the above discussion we have ignored terms $O(\ge ^2)$, and one may
be concerned about their relevance.
Fortunately,
the solution can be exactly found near the infalling line.
Following ref. [\rst ], we
consider the equations of motion along a light-like line infinitesimally
above the matter trajectory,
$x^+=1/\lambda $. On this line they
are ordinary differential equations, in the variable $x^-$, for the quantities
$\p_+ X$ and $\p_+ Y$.
If $X_0$ and $Y_0$ denote the linear dilaton vacuum solution, we have
$$
\p_-(\p_+ X)=A e^{Y_0}\ ,
\eqn\eqxxx
$$
$$
\p_-(\p_+ Y)=B e^{X_0} +Ce^{Y_0}\ .
\eqn\eqyyy
$$
By integrating over $x^-$ one finds
$$
\p_+ X=\mu e^{{4\lambda\over\kappa}x^-}\big[ {\kappa\over 4\lambda }-x^-\big]
\ ,
\eqn\dpx
$$
$$
\p_+Y={4\gl\over \kappa}(x^-+{m\over \gl})+\lambda+
{4\kappa\over\kappa-16}\mu e^{{4\lambda\over\kappa}x^-}\big[
{\kappa\over 4\lambda }-x^-\big]\ .
\eqn\dpy
$$
Hence
$$
\p_+\Omega =-{\gl\over\sk }(x^-+{m\over\gl})-{\sk\over 4}\lambda
-{\sk\over 4}{3\kappa-16\over \kappa-16}\mu e^{{4\lambda\over\kappa}x^-}
\big[ {\kappa\over 4\lambda }-x^-\big]\ .
\eqn\esdpO
$$
For $\kappa\sim 16$ eq. \esdpO \ reduces to eq. \plusomga\ {\it cum }
$x^+=1/\lambda$.
The generalization to arbitrary $\kappa$ of eq. \quemu\ is found
from eq. \esdpO :
$$
\hat\mu (\lambda ,\kappa ,m)=
-{8e(\kappa -16)\over \kappa^2(3\kappa-16)} m\lambda\ .
\eqn\musombrero
$$

\chapter{Hawking radiation}

Thus we see that non-perturbative corrections can easily modify
the causal character of the boundary line and hence the space-time
topology.
It is reasonable to expect that, with suitable boundary conditions,
the boundary curve will stay
time-like, asymptotically approaching some null line $x^-=-v $, with
$0<v < -x^-_0 $. The geometry is depicted in fig. 4, which resembles
the subcritical case discussed in sect. 2 (see fig. 3).
Let us assume that the system finally decays into the vacuum.
At $x^+>>1/\lambda $ the solution will take the form
$$
\chi=\Omega =-\gl x^+ (x^-+v)-\ln [-\gl x^+(x^-+v)]\ ,
\eqn\ldv
$$
or
$$
ds^2=-d\tau ^2+d\sigma ^2\ ,\ \ \ \phi=- \lambda \sigma\ ,
\eqn\lldv
$$
where
$$
e^{\lambda\sigma^+}=\lambda x^+\ ,\ \ \ \ e^{-\lambda \sigma^-}=-\lambda
(x^-+v)
\ ,\ \ \ \ \sigma^\pm=\tau\pm\sigma\ \ .
\eqn\mincoord
$$

The Hawking radiation can be computed in the standard way [1]
(for a derivation in the context of dilaton gravity see
ref. [\gidnel ]).
It is useful to introduce Minkowski coordinates for the region
$x^+<1/\lambda $:
$$
e^{\lambda y^+}=\lambda x^+\ ,\ \ \ \ e^{-\lambda y^-}=-\lambda x^-\ \ .
\eqn\coormin
$$
The mode expansions for the right moving field are
$$\eqalign{
f_-&=\int_0^\infty d\omega \ [a_\omega u_\omega+a^{\dag }_\omega u_\omega^*]
\ \ \ \ \ \ \ ({\rm in}) \ \ ,\cr
&=\int_0^\infty d\omega \ [b_\omega v_\omega+b^{\dag }_\omega v_\omega^*]
\ \ \ \ \ \ \ ({\rm out})\ \ ,\cr }
\eqn\modos
$$
where
$$\eqalign {
u_\omega &={1\over \sqrt{2\omega} }e^{-i\omega y^-}\ \ \ \ ({\rm in})\ \ ,\cr
v_\omega &={1\over \sqrt{2\omega} }e^{-i\omega \sigma^-}
\ \ \ \ ({\rm out})\ .\cr }
\eqn\base
$$
The in and out vacuum are defined by
$$
a_\omega|0\rangle _{\rm in}=0\ ,\ \ \ b_\omega|0\rangle _{\rm out}=0\ .
\eqn\vacios
$$
The calculation of the Bogoliubov coefficients is analogous to ref. [\gidnel ],
so we will not repeat it here. For the number operator for out modes,
$N_\omega^{\rm out}=b^{\dag }_\omega b_\omega $, one has
$$
{}_{\rm in}\langle 0|N_\omega^{\rm out}|0\rangle_{\rm in}=
\int_0^\infty d\omega' \ |\beta_{\omega\omega'} |^2\ ,
\eqn\opnumero
$$
with
$$
\beta_{\omega\omega'} ={1\over 2\pi\lambda }
\big({\omega '\over\omega-\epsilon }\big)^{1/2} (\lambda v)^{i\omega/\lambda}
B(u_1,u_2)\ ,\ \ \
\eqn\bogol
$$
$$u_1=-{i\over\lambda }(\omega'+\omega)+\epsilon\ ,\ \ \
u_2=1+{i\omega\over\lambda}\ .
$$
At late times this leads to a thermal distribution with temperature
$T_{\rm H}=\lambda/2\pi $ [\gidnel ], as is characteristic of two-dimensional
models [\witten].

The expectation value of the energy momentum tensor,
${}_{\rm in}\langle 0| T_{\mu\nu}^f |0\rangle _{\rm in}$,
asymptotically in the out region ${\cal J}_+$,
is computed in the standard way by normal ordering with respect
to $b_\omega,\ b_\omega^{\dag } $, i.e. one requires
${}_{\rm out}\langle 0| T_{\mu\nu}^f |0\rangle _{\rm out} =0$.
The result is
$$
{}_{\rm in}\langle 0| T_{--}^{\rm (i)} |0\rangle _{\rm in}=
{\kappa\over 4} \big[ {1\over (x^-+v)^2} - {1\over {x^-}^2}
\big]\ .
\eqn\temenos
$$
In region (ii) the precise form will depend on the boundary conditions.

Note that $T^{\rm (i)}_{--}$ vanishes for $x^-<<-v$,
$$
T^{\rm (i)}_{--}\sim -{\kappa\over 2} {v\over {x^-}^3}\ ,\ \
\ \ x^-<<-v\ .
$$
In particular, if $m>>\kappa\lambda$,
$T_{--}$ will be negligible at the fake event horizon at
$x^-\cong -m/\gl$. In fact, it is clear that most of the energy will
be radiated far beyond $x^-=-m/\gl$, in sharp contrast with the
usual picture of Hawking radiation.
At $x^-=-m/\gl$, the Bondi mass will be of the same order of the
total ADM energy carried by the shock wave. Indeed, the total energy radiated
out in region (i) is
$$\eqalign{
E_{\rm out}^{\rm (i)}&=-\lambda \int_{-\infty}^{x_0^-} dx^-
\big( x^-+v \big) T_{--}^{\rm (i)}  \cr
&=\gl v -{\kappa\lambda\over 4}\ln \big( 1- {4\lambda v\over \kappa }\big)\ ,
\cr}
\eqn\pocaroba
$$
which is a planckian order energy. The Hawking temperature is the same but
the radiation comes out at later times.
The boundary conditions will dictate how much of the total energy will
originate from pure reflection off the boundary, and how much of it
will be carried out as Hawking radiation.

In the usual semi-classical picture one assumes that the final
state has the form
$$
\chi=\Omega =-\gl x^+ (x^-+{p\over\gl})-\ln [-\gl x^+(x^-+{p\over \gl})]\ ,
\eqn\stavacuum
$$
with $p\cong m$. Quantum fluctuations of the endpoint position can only
correct $p$ by a planckian-order energy, and therefore eq. \stavacuum\
and the consequent spectrum of Hawking radiation
should not receive important corrections for macroscopic black holes.
However, we have just seen that, if the actual boundary of the
space-time is time-like, there is no way the final state can
have the form \stavacuum . It will be given by eq. \ldv , which
is different from \stavacuum\  in an important way, as far as
the information loss problem is concerned.

Given the geometry of fig. 4,
an observer who never crosses the null line $x^-=-{m\over\gl }$ will
undergo acceleration all time, approaching the speed of light as
$t\to\infty $. As a result, he will be immersed in a bath of thermal radiation,
detecting the same outgoing radiation that one would calculate if the
vacuum were given by eq. \stavacuum . The vacuum \ldv\ has a well-understood
physical meaning, i.e. the absence of particles according to all inertial
observers in the asymptotic region.

\chapter{Numerical analysis for specific models of black hole evaporation }

Let us consider a scenario in which the linear dilaton vacuum
receives a planckian order non-perturbative correction,
$\mu_{\rm ldv}=O(\gl )$. For definiteness let us take
(see eqs. \nombre\ and \musombrero )
$$
\mu =\hat\mu + \mu_{\rm ldv}=
-{8e(\kappa -16)\over \kappa^2(3\kappa-16)} m\lambda\ +\mu_{\rm ldv}\ ,
$$
$$
\mu_{\rm ldv}=\hat\mu (\lambda,\kappa ,m={\kappa\over 4}\lambda )\ ,
$$
The qualitative time-evolution of the geometry is independent
of $m$, as long as $m$ is much greater than the Planck mass. For
$m\sim {\kappa\over 4}\lambda $ some anomalous behaviour occurs
(in virtue of a collapsing of the apparent horizons), but
the present semi-classical approximation is not supposed to apply
for black holes of planckian mass.
So let us restrain our attention on macroscopic black holes.
A typical Kruskal diagram is exhibited in figs. 5 and 6.
These plots have been made with
$\lambda=1$, $\kappa\cong 16$ and $m=20$.
Many other cases of $\kappa >16$
and $m$ have also been investigated, in essence obtaining the same picture.

The geometry agrees with the standard semi-classical configuration
(see fig. 2) in weak-coupling regions which are not in causal
contact with strong coupling regions.
The inner apparent horizon starts on $x^+=1/\lambda $
at some $x^-$ near $x^-_0$ (fig. 6),  and it joins the outer apparent
horizon at the endpoint of the trapped region,  $x^+=x^+_{\rm e}$
(see appendix).
In addition there is another apparent horizon with
$\p_-\Omega=0$, but it entirely resides in a region where the
perturvative method to solve the differential equations is not very
reliable.
It is unclear whether this apparent horizon will subsist in the exact
solution.
In the case of fig. 5, it approaches asymptotically the null line
$x^-=-m/\gl $, but the approximation
breaks down much earlier, as indicated in the fig. 6.
The contours of constant $\Omega $ have
$${dx^+\over dx^-}=-{\p_-\Omega\over\p_+\Omega}\ ,$$
so they cross the apparent horizons with $\p_-\Omega=0$ and $\p_+\Omega=0$
with derivatives equal to zero and infinity, respectively.

There is a naked singularity at the time-like curve $\Omega=\Omega_{\rm cr}$,
so boundary conditions are needed for the continuation to region (ii).
This time the simplest choice eq. \bc \ cannot be implemented, as can be
easily verified.
Conceivably the boundary conditions are also corrected by
non-perturbative terms.
However, from eq. \curvature\ it seems clear that boundary conditions
which do not obey eq. \bc\ will necessarily lead to naked singularities,
presumably leading to instabilities. A black hole could evolve
into an object carrying an arbitrary amount of negative energy
and then continue to radiate {\it seculum secularis}.
This feature could be an artifact of the particular model
we have contemplated, or simply an artifact of the semi-classical
approximation.
Also, the shock-wave case is rather unphysical; it should be easier
to implement boundary conditions in the case of continuous distributions
of incoming matter, but in the present case this is
{\it terra incognita}, because there is no hint on what is
$\mu $; there are too many variables and it is hard to
account for all the possibilities.

Other possible extensions of the solution are suggested in fig. 7a and 7b.
In order to avoid a remnant scenario, the solution must approach the linear
dilaton vacuum for late times $x^+$. The necessary condition
is that, for
$x^-\leq x^-_0$ and $x^+>>1/\lambda $, $\Omega $ takes the form
\ldv \ . Unfortunately, the present approximation for solving the differential
equations \eqx ,\ \eqy\ breaks down as $\ge=O(1)$ (fig. 6).
At $x^-=x^-_0$ this corresponds to
$$
x^+\sim {m \over \lambda(m-{k\over 4}\lambda )}\ .
$$
So the exact solution is necessary to decide
whether the matching with the linear dilaton vacuum
on $x^-=x^-_0$ and $x^+>>1/\lambda $ is feasible for this specific model.

Without a cognition of the asymptotic
behaviour of the solution in ${\cal J}_+$ and in
region (ii) it is not
possible to determine the outgoing spectrum of Hawking radiation.
For physical reasons, it is likely that the boundary curve will stay time-like,
and it is conceivable that the system will finally decay into the vacuum.
Then the discussion of sect. 4 will apply and, in particular,
the outgoing energy-momentum tensor will be given by eq. \temenos .
Meanwhile, it is interesting to look at local quantities which in
certain limits are related to the Bondi mass of the black hole, for example,
the value of $e^{-2\phi}$ at the outer apparent horizon.
In the case of fig. 5, $m=20$, $\lambda=1$, $\kappa\sim 16$, one finds by
numerical computation that
$$
e^{-2\phi}|_{x^+=x^+_{\rm e}}=16.6\cong {m\over\lambda }-{\kappa\over 4}
\ .
\eqn\mbon
$$
This is unlike the $\mu=0$ RST case, where $e^{-2\phi}$ at the
endpoint is of planckian order. The apparent horizon deviates
from the $\mu=0 $ apparent horizon at earlier times than expected.
We have verified that this feature is
independent of any particular choice of the parameters, i.e.,
for macroscopic black holes $e^{-2\phi}$ at $x^+_{\rm e}$ seems to be,
roughly,
of the same order as $m/\lambda $, which suggests that most of the
energy will be radiated far beyond $x^-=-m/\gl $.
For example, for $m=40$ and $m=80$ one finds, respectively,
$e^{-2\phi}|_{x^+=x^+_{\rm e}}=21.5$ and
$e^{-2\phi}|_{x^+=x^+_{\rm e}}=24.8$. Unfortunately, there are
numerical problems to study the case of larger masses.
While the non-perturbative term we added to the action is
insignificant at the endpoint,
the fields $\Omega$ and $\chi $, being integrals of this, receive
non-negligible corrections near the endpoint.
Indeed, to the leading order in perturbation theory we
are making, $\Omega$ and $\chi $ contain terms of the form
$\mu \ge /(x^-+m) $.
For a very massive black
hole, $\ge $ is exponentially small at the endpoint of the trapped region
but $x^-$ is exponentially close to $-m$, giving rise to a contribution
of order unity. This explains why there seem to be some changes in local
quantities at the endpoint of the trapped region.
However, one must be careful in extracting
conclusions from this, since the present leading order approximation could
be simply breaking down before getting to the endpoint.
Certainly, it would be very interesting to have the exact solution to the
differential equations \eqx \ and \eqy .

\bigskip
\chapter{No cosmological veil conjecture}

{\rightline {\it Non lasciate ogni speranza, voi ch'entrate}}
\bigskip

In sects. 3 and 5 we have seen  simple models where additional
quantum corrections turn the space-like boundary into a time-like boundary,
altering the topology of the standard semi-classical picture of black hole
evaporation. In these models there is no longer a clear problem of
information loss, since all information may easily return by simply reflecting
back on the boundary curve.

It is tempting to speculate that the very nature of a quantum theory
is incompatible with space-like boundaries, and thus any illusory
space-like boundary of a semi-classical analysis is `washed out'
or `impelled' to a time-like boundary when the full quantum theory is
taken into account. Intuitively, quantum fluctuations would penetrate and
destroy any space-like boundary they may encounter in their future, averting
the formation of global event horizons.
A similar phenomenon should occur in other cases of topology change.

As far as the resolution of the information problem is concerned,
the topology does not
need to be trivial. For example, there could be a conical singularity
at the endpoint, as indicated in fig. 8. A large whormhole
would carry all the information back in region (ii).

The no veil conjecture may be formulated in a simple way:

{\it Quantum gravity precludes the formation of global
event horizons.}

That a global event horizon cannot be a strict `point of no return'
in  a quantum theory is obvious, since in quantum mechanics it
is not possible to localize, e.g., the endpoint or any branching
point with an infinite accuracy.
However, for large black holes, the fluctuations in the position
of the event horizon can be neglected compared to the Schwarzchild
radius. The above conjecture affirms that there are no global event
horizons, not even in an approximative sense.

Let us consider a black hole-type configuration with mass much larger than
the Planck mass. What will an outside observer see?.
Freely-falling matter will pass through the outer apparent horizon, then
enter into a strong coupling region, experiencing planckian curvatures,
and eventually will reflect back at zero radial coordinate.
An outside, time-like observer, far away from the black hole, will detect an
insignificant flux of Hawking radiation coming out
(as described in sect. 4). At time $x^-=-m/\gl $
he will not measure anything particular, the curvature will be almost zero
and the Hawking radiation will still be weak.
Finally, he will be in causal contact with the time-like boundary,
recovering all the energy and the quantum mechanical information,
including global quantum numbers (unless boundary interactions violate
the corresponding global symmetry).


\chapter{Critique }

Without pretense of a deep enquiry at this primitive stage,
it may be worth to mention a number of points which arise skepticism.
To begin with,
the examples investigated in this paper, though introduced only
for illustrative purposes,  are rather {\it ad hoc} and
represent an oversimplified sample of non-perturbative corrections. The
mere inclusion of other
terms in the potential \poten\ could modify the picture,
maybe in a favorable way,  but maybe unfavorably.
The fact that the null line $x^-=-m/\gl $ is still special in the
approximate solution is not really worrying, since this may be
an artifact of the perturbation theory around
$\Omega_{\rm class}=-\gl x^+(x^-+m)+...$ .
 Rather, what is more concerning is the issue of the boundary conditions.
This is important in order to define a setting for the construction of an
$S$-matrix. For stability reasons,
one would like to demand eqs \bc \ on the boundary so as
to ensure finite curvature everywhere.
It is necessary to have a time-like line starting
at $\phi=\phi_{\rm cr}$ with $\p_+\Omega =0$.
In the numerical study there was no way to achieve this,
irrespective of the choice of parameters, which jeopardizes the
implementation of the boundary conditions. It would be interesting
to show that, e.g. by adjusting different parameters of the potential,
one can have the structure of fig. 4 and a time-like line starting
at $\phi=\phi_{\rm cr}$ with $\p_+\Omega =0$ (which shall be
the boundary curve after implementation of boundary conditions).
A last recourse is invoking new degrees of freedom for the
space-time boundary.
A related point is that, even in the specific examples provided in sect. 5,
there is not enough evidence to believe that the  geometry
will eventually decay into the vacuum. It might approach some
static solution, producing a remnant scenario.

However, with no intention of stating a theorem, we have seen that
non-perturbative corrections which are only important in strong coupling
regions may lead to a picture which is completely different
from the usual semiclassical picture.
This is somewhat surprising, since in standard two-dimensional
dilaton gravity there are local quantities that can be associated with the
Bondi mass. Given the vacuum \ldv ,
this correspondence will not be correct for inertial observers,
 as follows from the discussion of sect. 4,
whereas it will still hold for non-inertial
observers that never cross the line $x^-=-m/\gl $.
But, as well-known, Hawking radiation is a global effect and it
is no sense associated with local physics in the vicinity of the horizon.
Perhaps the lesson we learn from these exercises is that
the standard model of black hole evaporation by emission
of Hawking radiation is in a firm ground only if one presupposes that
the topology of the space-time is not modified at quantum level.
Barring this assumption, Hawking's prediction is delicate
and pendent on further quantum corrections like a Damocles sword.

Leaving aside the two-dimensional example and its problems,
the proposal for the resolution of
information problem maintains a conservative viewpoint in the sense it does
not lead to violation of the standard laws of quantum mechanics and
thermodynamics, and in weak coupling regions,
which are not in the causal future of strong coupling regions,
it locally agrees with the usual semiclassical picture.
It is somewhat radical in the sense that
it resolves the information problem by removing from the stage
the very origin of the paradox, the black holes.
Nevertheless, it is not at all clear why there should always be
an inner apparent
horizon (or an odd number of apparent horizons),
 insensitive to distinct cases of Cauchy data, which would
permit the space-time boundary to align in a time-like direction.
It is suspicious that this should occur without pronouncing new laws
of physics, just as a consequence of ``standard laws of quantum mechanics".
The appearance of a space-like boundary does not violate any quantum
mechanical axiom, there is nothing inherent to quantum mechanics
locally incompatible with space-like boundaries,
and therefore it is unclear why the ``no cosmological
veil conjecture" should be true.

Regrettably, the alternatives, called 1) and 2) in the introduction,
are practically ruled out for reasons which seem to be insurmountable.
Lacking enough justification in support of the no veil conjecture,
this remains a theoretical caprice.
Clearly, there is much work to be done.

\bigskip\bigskip
 \noindent $\underline {\rm Acknowledgements}$: The author wishes to thank
J. Polchinski, L. Susskind and L. Thorlacius for valuable discussions.

\vfill\eject

\centerline {\bf Appendix}
\bigskip
{\captura APPARENT HORIZONS IN TWO DIMENSIONS}

The apparent horizon plays an important role in the mechanisms
described in the main text, and it is also a very useful and
physically meaningful object in the standard picture of black hole
formation and evaporation. Thus it is worth to refresh the connection
with the standard four dimensional definition (see e.g. ref. [\wald ]).

Let $C$ be a three-dimensional manifold with boundary $S$.
Let $\xi _\mu$, $\mu=0,1,2,3$, be the vector field of tangents to a
congruence of
outgoing null geodesics orthogonal to $S$. $C$ is a {\it trapped region}
if the expansion $\theta=\nabla_\mu\xi^\mu $ is everywhere non positive
on $S$, $\theta \leq 0$. The {\it apparent horizon} ${\cal A}$ is the
boundary of the {\it total trapped region}, the latter defined as the closure
of the union of all trapped regions. A corollary of this definition
is that $\theta=0$ on ${\cal A}$.

Now let us contemplate metrics of the form
$$
ds^2=g_{ij}(x^0,x^1) dx^idx^j+ \exp [-2\phi (x^0,x^1)] d\Omega ^2
\ \ ,\eqno (A.1)$$
where $i,j=0,1$. In this spherically symmetric space-time we have
$\xi_\mu=\{ \xi_0,\xi_1,0,0 \}$, and the geodesic equation reduces to
$$
\xi^i\nabla_i\xi^j =0\ \ ,
\eqno(A.2)
$$
i.e. the two-dimensional geodesic equation. Since in this dimensionally
reduced configuration there is only one family of outgoing
null geodesics, a trapped region is the total trapped region, and the
condition determining the apparent horizon simply becomes
$$\theta=0\ . \eqno(A.3)$$
{}From eq. (A.1) one easily obtains
$$
\theta =\theta ^{(2)} -2\xi^j\p_j\phi\ \ ,
\eqno(A.4)
$$
where
$$
\theta ^{(2)} \equiv \p_i\xi^i+\Gamma_{ij}^i\xi^j\ \ .
\eqno(A.5)
$$

Let us denote $B_{ik}=\nabla _i\xi_k $. By using the geodesic equation,
$\xi^iB_{ik}=0$,
and the fact that $\xi $ is null, $\xi^i\xi_i=0$, one derives the following
relations
$$
\xi^1B_{11}=-\xi^2B_{21}\ ,\ \ \ \xi^1B_{21}=-\xi^2B_{22}\ ,\ \ \ B_{12}=B_{21}
\ \ ,
\eqno(A.6)
$$
thereby we obtain
$$
\eqalign{
\theta^{(2)} &=g^{ij}B_{ij}=B_{11}\big[ g^{11}-2{\xi^1\over\xi^2}g^{12}
+\big({\xi^1\over\xi^2}\big) ^2g^{22}\big]  \cr
&=0 \ ,\cr}
\eqno(A.7)
$$
where we have used $\xi^1\xi_1=-\xi^2\xi_2$. Thus we see that the
two-dimensional expansion parameter is identically zero. This means
that an intrinsically two-dimensional apparent horizon can not be defined.
Now, by using eqs. (A.3), (A.4) and (A.7) we find that the
condition defining the
apparent horizon becomes
$$
\xi^i\p_i\phi = 0
\eqno(A.8)
$$
Since $\xi $ is null eq. (A.8) implies $\xi_i = f(x) \p_i\phi $, where
$f(x)$ is a function. Therefore the condition (A.3) translates to
$$
g^{ij}\p_i\phi\p_j\phi=0
\eqno(A.9)
$$

In the conformal gauge eq. (A.9) reduces to
$$
\p_+\phi\p_-\phi=0
\eqno(A.10)
$$
Therefore the apparent horizon ${\cal A}$ is the locus of
$ \p_+\phi \p_-\phi $. In terms of $\Omega(\phi )$,
eq. (A.10) reads
$$
{1\over \Omega '{}^2} \p_+ \Omega \p_-\Omega =0
\eqno(A.11)
$$
that is, provided $\Omega'\neq 0$, the points satisfying
$\p_+\Omega =0$ or $\p_-\Omega =0$ define the position
of the apparent horizon. In the critical line one may have
$\p_+\Omega=\p_-\Omega=0$ but $\p_+\phi\p_-\phi\neq 0$, as
occurs in the subcritical case of sect. 2 when the
boundary conditions \bc\ are applied.

\refout
\vfill\eject
\null
\centerline {\bf Figure Captions}

\bigskip

\n {Fig. 1a: Penrose diagram corresponding to the standard semi-classical
picture of black hole evaporation.}

\bigskip

\n {Fig. 1b: Penrose diagram of another geometry which differs
from fig. 2a only in the strong curvature region. Here there is no space-like
boundary.}

\bigskip

\n {Fig. 2: Standard semi-classical picture of black hole evaporation
in Kruskal coordinates, corresponding to gravitational collapse
of an incoming shock wave ($\lambda=1$).}

\bigskip

\n {Fig. 3: A subcritical incoming energy flux leads to a time-like
singularity.}

\bigskip

\n {Fig. 4: Qualitative picture of black hole formation and
evaporation in the model of sect. 3.}

\bigskip

\n {Fig. 5: Geometry corresponding to the model of sect. 5, illustrated
by numerical plots of contours of constant $\phi $.
The dashed line indicates
the region where the present approximation begins to break down.
Figures a), b) and c) correspond to different regions and scales of
the same configuration.}

\bigskip

\n {Fig. 6: Numerical plot of the apparent horizons in the geometry of
fig. 5.}

\bigskip

\n {Fig. 7: Possible extensions of the geometry of figs. 5
into the region where the solution to the differential equations
\eqx , \eqy\ is unknown. a) The apparent horizon with $\p_-\Omega=0$
is absent; b) The apparent horizon with $\p_-\Omega=0$ is present,
but it is closed, confined to a finite region in space-time.
}

\bigskip

\n {Fig. 8: An alternative topology instead of fig. 1b which would
lead to similar results.}

\bigskip

\vfill\eject
\end